\documentclass[preprint,11pt]{article}
\usepackage{amssymb,amsfonts,amsmath,amsthm,amscd}
\usepackage{graphicx}% Include figure files
\usepackage{epsfig}% Include figure files

\newtheorem{df}{Definition}
\newtheorem{thm}{Theorem}

\newtheorem{lemma}{Lemma}

\def\gt{\tilde{g}}
\def\Z{Z_N}
\def\ve{\varepsilon}
\def\da{{\partial a}}

\def\dpi{{\partial_{+} i}}
\def\dpj{{\partial_{+} j}}
\def\dmi{{\partial_- i}}
\def\dmj{{\partial_- j}}
\def\ed{\stackrel{\rm d}{=}}

\def\uh{\underline{h}}
\def\oh{\overline{h}}
\def\uu{\underline{u}}
\def\ou{\overline{u}}

\def\D{{\mathcal D}}
\def\R{{\mathbb R}}
\def\E{{\mathbb E}}
\def\prob{{\mathbb P}}

\def\ux{\underline{x}}
\def\uy{\underline{y}}
\def\uz{\underline{z}}

\def\sBP{{\rm {\tiny BP} }}
\def\sTV{{\rm {\tiny TV} }}

\def\l|{\left|\left|}
\def\r|{\right|\right|}

\def\Ball{{\sf B}}
\def\T{{\sf T}}
\def\hT{\widehat{\sf T}}
\def\poisson{{\sf Poisson}}
\def\ind{{\mathbb I}}

\def\prooft{\hspace{0.5cm}{\bf Proof:}\hspace{0.1cm}}
\def\endproof{\hfill$\Box$\vspace{0.4cm}}

\newcommand{\<}{\langle}
\renewcommand{\>}{\rangle}

\begin{document}

\title{Counting good truth assignments of random $k$-SAT formulae}

\author{Andrea Montanari\thanks{Laboratoire de Physique 
Th\'{e}orique de l'Ecole Normale Sup\'{e}rieure, Paris.
Research is partially supported by European Union under the ip EVERGROW. Email: {\tt montanar@lpt.ens.fr}} \;\; and\; 
Devavrat Shah\thanks{LIDS, MIT. Research is partially supported by NSF CAREER. 
Email: {\tt devavrat@mit.edu}.
}}

\date{\today}
\maketitle

\thispagestyle{empty}

\abstract{We present  a deterministic approximation algorithm to compute
\emph{logarithm} of the number of `good' truth assignments for
a random $k$-satisfiability ($k$-SAT) formula in polynomial time
(by `good' we mean that violate a small fraction of clauses). The 
relative error is bounded above by an arbitrarily small constant
$\epsilon$ with high probability\footnote{In this paper, by term "with high probability" (whp) we
mean with probability $1-o_N(1)$.} as long as the clause density (ratio of
clauses to variables) $\alpha<\alpha_{\rm u}(k) = 2k^{-1}\log k(1+o(1))$.
The algorithm is based on computation of marginal distribution via belief 
propagation and use of an interpolation procedure. This scheme 
substitutes the traditional one based on approximation  of
marginal probabilities via MCMC, in conjunction with self-reduction,
which is not easy to extend to the present problem. 

We derive $2k^{-1}\log k (1+o(1))$ as threshold for uniqueness of the 
Gibbs distribution on satisfying assignment of random infinite tree 
$k$-SAT formulae to establish our results, which is of interest in its own right. 
}
%
%*******************************************************************
%
\section{Introduction}

\noindent{\bf Setup and Problem Statement.} 
Given $N$ boolean variables $x_i, 1\leq i\leq N$, an $M$ clause
$k$-satisfiability ($k$-SAT) formula has the form $F = \wedge_{j=1}^M C_j$, 
where  $C_j = \vee_{\ell=1}^k z_{j_\ell}$ with literal $z_{j_\ell}$ being 
either $x_i$ for $\bar{x}_i$ for some $1\leq i\leq N$. An assignment 
$\ux \in \{0,1\}^N$
of variables $x_i, 1\leq i\leq N$ satisfies clauses $C_j$ if at least of
one the $k$ literals of $C_j$ evaluates to be true. We will denote 
true by ``1'' and false by ``0''.  For given $F$, $E(\ux)$ denote 
the number of unsatisfied clauses of $F$ under assignment $\ux$.
Given $\beta \in {\mathbb R}_+$ (called {\em inverse temperature} in 
statistical physics), define {\em partition function} as 
\begin{eqnarray}
\Z(\beta,F)  \equiv \sum_{\ux \in \{0,1\}^N}e^{-\beta E(\ux)}\, .\label{eq:ZDefinition}
\end{eqnarray}
Notice that $\Z(\beta, F)$ weighs in favor of ``good'' assignments, 
i.e. assignments that satisfy more clauses. As $\beta \to \infty$, 
$\Z(\beta, F)$ becomes the number of assignments that satisfy
(all clauses of) $F$.  
The partition function naturally arises as normalizing constant in
the following probability measure on $\{0,1\}^N$, often denoted as 
{\em Boltzmann}
distribution \cite{Georgii} related to $F$: for $\ux \in \{0,1\}^N$,
\begin{eqnarray}
\mu_{\beta, F}(\ux) = \frac{1}{\Z(\beta, F)}\prod_{j=1}^M\psi_j(\ux) ~~=~~ \frac{e^{-\beta E(\ux)}}{\Z(\beta, F)} \, ,\;\;\;\;\mbox{where}
\;\;\;\;\;\psi_j(\ux) = \left\{\begin{array}{ll}
1 & \mbox{ if $\ux$ satisfy clause $C_j$,}\\
e^{-\beta} & \mbox{ otherwise.}
\end{array}\right.\label{eq:GraphicalModel}
\end{eqnarray}
We shall write $\mu(\,\cdot\,) = 
\mu_{\beta,F}(\,\cdot\,)$  whenever it will not be necessary
to specify the formula and inverse temperature. 
We further denote by $\<\,\cdot\,\> =\<\,\cdot\,\>_{\beta,F}$ 
expectations with respect to the measure $\mu$. 

In this paper, we are interested in {\em random $k$-SAT} formulas. 
These are generated by
selecting $M$ clauses independently and uniformly at random from all possible
$2^k {N \choose k}$ $k$-clauses. Specifically, let $M$ scale linearly in $N$, i.e. 
$M = \alpha N$ for $\alpha \in {\mathbb R}_+$.

The main motivation in this paper is to describe an
efficient algorithm to compute a good approximation of $\Z(\beta, F)$ 
for such random formulas. 
An important open conjecture is to show, that for any 
$\alpha, \beta \in {\mathbb R}_+$, 
under the probability distribution induced by random $k$-SAT formula, 
the limit $\lim_{N\to\infty} \frac{1}{N} \log \Z(\beta, F)$ 
exists with probability $1$. The analysis of our algorithm implies
such a result for all finite $\beta$, and $\alpha$ smaller than a critical 
value.

\vspace{.1in}
\noindent{\bf Related Previous Work.} The well-known threshold conjecture 
for random $k$-SAT states that for all $k \ge 2$, there exists 
$\alpha_{\rm c}(k)$ such
that for $\alpha < \alpha_{\rm c}(k)$ (resp. $\alpha > \alpha_{\rm c}(k)$) 
the randomly generated formula is satisfiable (resp. not satisfiable) with 
probability $1$ as $N\to\infty$.  There has been a
lot of interesting work on this topic, and a convergence of
methods from different communities \cite{Monasson,Mezard,ANPNature}. 
Due to space limitation, we will recall only some of the key relevant results.

Friedgut \cite{Friedgut} established existence of a sharp threshold.
More precisely, he proved that  there exists 
$\alpha_{\rm c}(k, N)$ such that the satisfiability probability
tends to $1$ (to $0$) if $\alpha<\alpha_{\rm c}(k,N)(1-\eta)$
(respectively $\alpha>\alpha_{\rm c}(k,N)(1+\eta)$). 
While it is expected that $\lim_{N\to\infty} \alpha_c(k,N)$ exists,
it has still remained elusive. Recently, Achlioptas and Peres \cite{AP04} 
established  that $\alpha_{\rm c}(k, N) = 2^k \ln k (1+o_k(1))$ 
thus implying that  $\alpha_{\rm c}(k, N)$ can be taken $N$ independent 
to first order for large $k$. 

The existence of $\lim_{N\to\infty} \lim_{\beta\to\infty} 
\frac{1}{N} \log \Z(\beta, F)$ with probability $1$, for all 
$\alpha \in {\mathbb R}_+$ and $k$ naturally establishes the threshold 
conjecture. 
More generally, the log-partition function at $\beta=\infty$ 
provides detailed information about the satisfying assignments 
(computing it exactly is of course $\#$-P complete). 
In \cite{MonassonZecchina} a formula for the limit log-partition
function was derived through the non-rigorous replica method
from statistical physics.
The existence of the $N\to\infty$ limit was proved by 
Franz, Leone and Toninelli \cite{FL03, FLT03} for  even $k$ and  all values of 
$\alpha$. These authors also provided an upper bound on 
$\lim_{N\to\infty}\frac{1}{N} \log \Z(\beta, F)$. 
However evaluating the bound requires solving an {\em a priori} complex 
optimization problem, and a matching lower bound wasn't proved there.
Talagrand \cite{T01} established the existence 
of the limit and its value for very small value of  $\beta$ (depending 
on $k$). 

\vspace{.1in}
\noindent{\bf Overview of Results.} In this paper, we essentially prove that 
the Boltzmann distribution (\ref{eq:GraphicalModel}) is a
{\em pure state} \cite{Georgii} by establishing appropriate worst-case 
{\em correlation decay} for tree formulae. The approach of Talagrand \cite{T01}
also crucially relied of proving correlation decay, 
albeit with different means. This resulted in a limitation to small values of 
$\beta$ and thus leaving out interesting regime of large $\beta$.

An analogy can be drawn with the Markov Chain Monte Carlo (MCMC) approach
to the approximate computation of partition functions 
(see, for example, work by Jerrum and Sinclair \cite{JS93}).
In that case, the crucial step consists in proving an appropriate
mixing condition (`temporal' correlation decay) for some Markov Chain.
The same role is played here by `spatial' correlation decay with respect 
to the measure (\ref{eq:GraphicalModel}).

In this paper, we establish correlation decay for random $k$-SAT formula for a
range of $\alpha$ and all $\beta$.
This allows to estabilish that deterministic Belief Propagation algorithm
provides a good approximation of the marginals with respect to
the distribution (\ref{eq:GraphicalModel}), cf. Section \ref{sec2}.
In the usual MCMC approach, marginals are used to approximate the partition 
function by recursively fixing the variables and exploiting self-reducibility.
This cannnot be done in the present case because the reduced SAT
formulae are not random anymore.
Instead, we use  {\em interpolation} in $\beta$, to obtain 
$\log \Z(\beta, F)$ approximately (Theorem \ref{thm:KSAT}).
The analysis of the approximation scheme implies the existence of the limit 
$\lim_{N\to\infty}\frac{1}{N} \log \Z(\beta, F)$ 
(Theorem \ref{thm:LimitPartFun}).   
We hope that our novel approach for counting will find applications in 
other hard combinatorial problems. 
Similar  schemes were recently discussed by  Weitz \cite{W06}, and
Bandyopadhyay and 
Gamarnik \cite{BG06} for counting independent  sets approximately 
via deterministic algorithms.

Finally, we show that the computation of the partition function
leads to an estimate of the number of truth assignments that violate at most
$N\ve$ clauses, for small $\ve$  (Theorem \ref{thm:AlmostSAT}).
As a byproduct, we obtain an asymptotically (in $k$) threshold for
uniqueness Gibbs measure on infinite $k$-SAT tree formula
(Theorem \ref{thm:UniquenessTrees}).

\vspace{.1in}

\noindent{\bf Organization.} Section \ref{sec1} presents preliminaries and
statements of the main results. The Section \ref{sec2} describes the
approximate counting algorithm and  the proof of key Lemmas related 
to the correlation decay (or uniqueness) of Gibbs distribution on random 
{\em tree} $k$-SAT. The Section \ref{sec3} completes the proofs of all 
main results stated in Section \ref{sec1}.  We present direction for future work in 
Section \ref{sec4}. 

\section{Preliminaries and Main Results}
\label{sec1}

Given $\alpha$ and $k$, define $\alpha_*(k)$ to be 
the smallest positive  root of the equation $\kappa(\alpha) = 1$, 
where
\begin{eqnarray}
\kappa(\alpha) \equiv k(k-1)\alpha\left(1-\frac{1}{4}\,e^{-k\alpha/2}\right)
\left(1-\frac{1}{2}\,e^{-k\alpha/2}\right)^{k-2} \, .
\label{eq:ContractionRate}
\end{eqnarray}
For $k=2, 3, 4, 6$, the $\alpha_*(k)$ is approximately 
$0.58216, 0.293, 0.217, 0.16670$. Asymptotically, 
$\alpha_*(k) = 2k^{-1}\log k \left(1+ O\left(\frac{\log \log k}{\log k}\right)\right)$.
Now, we state the main result of this paper about approximating logarithm of
partition function. 
\begin{thm}\label{thm:KSAT}
Given $\ve > 0$ and $\alpha < \alpha_*(k)$, there exists  $\delta'>0$ and
a polynomial (in $N$,  independent of $\ve$) time algorithm that computes 
a number $\Phi(\beta,F)$ (the input being $\beta\in {\mathbb R}$ and a 
satisfiability formula $F$) such that the following is true.
If $\beta \in [0, N^{\delta'}]$ and $F$ is 
random $k$-SAT formula with $N$ variables and $M = N\alpha$ clauses,
then, with high probability,
\begin{eqnarray}
\Phi(\beta,F)\,(1-\ve)\le\log \Z(\beta,F) \le \Phi(\beta,F)\,(1+\ve). \,
\label{eq:Guarantee}
\end{eqnarray}
\end{thm}
The proof of Theorem \ref{thm:KSAT} requires us to prove uniqueness of Gibbs
measure for the model (\ref{eq:GraphicalModel}) on  infinite tree random $k$-SAT formulae. To state this result, we
first need some definitions. An appropriate model for tree random $k$-SAT, 
$\T_*(r)$ is described as follows: For $r=0$,  it is the graph containing 
a unique variable node. For any $r \ge 1$, start by a single variable node 
(the root) and add $l\ed \poisson(k\alpha)$ clauses, each one 
including the root, and $k-1$ new variables (first generation variables). 
For each one of the $l$ clauses, the corresponding literals are
non-negated or negated indipendently with equal probability. If $r \ge 2$, 
generate an independent copy of $\T_*(r-1)$ for each variable node
in the first generation and attach it to them. By construction, 
for any $r'<r$ the first $r'$ generations of a tree from
$\T_*(r)$ are distributed according to the model $\T_*(r')$.
As a consequence, the infinite tree distribution $\T_*(\infty)$ 
is also well defined. In what follows, we denote
the root of $\T_*(\cdot)$ as $0$.  Let $\mu$ denote the Gibbs distribution
on random formula on $\T_*(r)$ (cf. (\ref{eq:GraphicalModel})) and
$\mu_{0|r}(x_0|\ux_r)$ be the conditional distribution of root 
variable conditional to the assignment of $r$-th generation
nodes of $\T_*(r)$ according to $\ux_r$. The key property for most of
the results of this paper  is that of correlation decay with respect to
random tree formulas $\T_*(\cdot)$.
\begin{df}\label{def:Uniqueness}
Given $\alpha, \beta \in {\mathbb R}_+$ and $k\ge 2$, the Gibbs distribution
defined by (\ref{eq:GraphicalModel}) on the random tree $\T_*(\cdot)$
is unique with exponential correlation decay if there exists positive constants $A,\gamma > 0$,
such that
\begin{eqnarray}
\E\left[\sup_{\ux_r,\uz_r}
\l|\mu_{0|r}(\,\cdot\,|\ux_r)-\mu_{0|r}(\,\cdot\,|\uz_r)\r|_{\sTV} \right]
\le A\, e^{-\gamma r}\, ,
\end{eqnarray}
for any $r\ge 0$. The uniqueness threshold $\alpha_{\rm u}(k)$ is the supremum value 
of $\alpha$ such that the above condition is verified for any 
$\beta\in[0,\infty]$.
\end{df}
The property defined here is a lot stronger than the usual notion of 
correlation decay, which only requires 
$\l|\mu_{0|r}(\,\cdot\,|\ux_r)-\mu_{0|r}(\,\cdot\,|\uz_r)\r|_{\sTV}\to 0$ as $r\to\infty$
almost surely. Let $\alpha_{\rm u}'(k)$ denote the threshold for this
weaker property. To the best of our knowledge, nothing has been known
about the precise values of $\alpha_{\rm u}(k), \alpha_{\rm u}'(k)$ or the
relation between them other than trivial lower bound from percolation
threshold of $\Omega(k^{-2})$. We establish the precise asymptotic behavior
of $\alpha_{\rm u}(k)$ and show that $\alpha_{\rm u}(k) = 
\alpha_{\rm u}'(k)(1+o_k(1))$
as stated below.
\begin{thm}\label{thm:UniquenessTrees}
For the Gibbs distribution (\ref{eq:GraphicalModel}) defined on $\T_*(\cdot)$ as above,
\begin{eqnarray}
\alpha_{\rm u}(k) = \frac{2\log k}{k}
\left\{1+O\left(\frac{\log\log k}{\log k}\right)\right\}, \;\;\;\;\;\;\;\;
 \alpha_{\rm u}'(k) = \frac{2\log k}{k}
\left\{1+O\left(\frac{\log\log k}{\log k}\right)\right\}\, .
\end{eqnarray}
\end{thm}
Though algorithmically we obtain approximation of $\log \Z(\beta, F)$,  
it is possible to establish the convergence of $\frac{1}{N} \log \Z(\beta, F)$
with probability $1$. Before stating this result, we need some definitions. 
In what follows, define function $f:\R^{k-1}\to\R$ as
\begin{eqnarray}
f(x_1,\dots,x_{k-1}) = -\frac{1}{2}\log\left\{1-
\frac{1-e^{-\beta}}{2^{k-1}}\prod_{i=1}^{k-1}(1-\tanh x_i)\right\}\, .
\label{eq:FDefinition}
\end{eqnarray}
Let $\D$ denote the space of probability distributions on the 
real line $\R$. Define functions 
$S, S_1, S_2: \D \to \D$ as follows: Given $\mu \in \D$, define 
random variable $u = f(h_1,\dots, h_{k-1})$ where $h_1,\dots, h_{k-1}$ are 
i.i.d. with distribution $\mu$. Define distribution of $u$ as $S_1(\mu)$. 
Given a distribution $\nu \in \D$, let random variable $h_0 = \sum_{a=1}^{\ell^+} u_a - \sum_{b=1}^{\ell^-} u_b$,
where $\ell^+, \ell^-$ are independent Poisson random variables with
mean $k\alpha/2$ and $u_a, u_b$ be i.i.d. with distribution $\nu$. Let distribution
of $h_0$ be denoted by $S_2(\nu)$. Define  $S \equiv S_1 \circ S_2$.  Now,
we state the result.
\begin{thm}\label{thm:LimitPartFun}
Given $k$, let $\alpha< \alpha_*(k)$ and $\beta\in [0,\infty)$. Then, 
the function $S : \D \to \D$ as defined above has unique fixed point, say
$\mu^*$. Let $\nu^* = S_2(\mu^*)$. Then, 
\begin{eqnarray}
\frac{1}{N}\log Z(\beta,F_N)\stackrel{{\rm a.s.}}{\to} \phi(\beta)\, ,\label{eq:AlmostSure}
\end{eqnarray}
\begin{eqnarray}
\mbox{where}~\phi(\beta) &=& -k\alpha\E\log[1+\tanh h\tanh u]+
\alpha\E\log \left\{1-\frac{1}{2^{k}}(1-e^{-\beta})
\prod_{i=1}^{k}(1-\tanh h_i)\right\}
+\label{eq:phi}\\
&&+\E\log\left\{\prod_{i=1}^{\ell_+}(1+\tanh u^+_i)
\prod_{i=1}^{\ell_-}(1-\tanh u^-_i)+\prod_{i=1}^{\ell_+}(1-\tanh u^+_i)
\prod_{i=1}^{\ell_-}(1+\tanh u^-_i)\right\}\, ,\nonumber
\end{eqnarray}
where $u, u^{\pm}_i$ are i.i.d. with distribution $\mu^*$, $h, h_j$ are i.i.d.
with distribution $\nu^*$ and $\ell_{\pm}$ are Poisson of mean $k\alpha/2$. 

\end{thm}
Finally, define $\Xi(\zeta,F)$ to be the number of assigments that violate at
most $\zeta$ clauses. The next result formalizes the relation between the
approximation of $\Z(\beta, F)$ and counting the number of truth assignments
that violate a small fraction of clauses.
\begin{thm}\label{thm:AlmostSAT}
For any $k\ge 2$, $\ve > 0$, and $\alpha<\alpha_*(k)$
there exists $A,C>0$, $a>0$ such that the following is true. 
If $F$ is a  random $k$-SAT $M=N\alpha$ clauses over $N$ variables, 
and $\beta = A\log 1/\ve$, then
\begin{eqnarray}
\left|\log \Xi(N\ve,F)-\Phi(\beta,F)\right|\le NC\ve^a\, ,
\end{eqnarray}
with high probability, where $\Phi(\beta, F)$ as defined in Theorem 
\ref{thm:KSAT}.
\end{thm}

\section{Algorithm and Key Lemmas}
\label{sec2}

\subsection{Algorithm}
\label{ssec:algorithm}

We first define a factor graph $G_F$ for a given formula $F$: each
variable is represented by (circle) variable node and each clause
by a (square) clause node with an edge between a variable and
a clause node only if corresponding variable belongs to the clause. The
edge is solid if variable is non-negated and dashed if variable is
negated. The Belief Propagation (BP) algorithm is a heuristic (exact for
tree factor graphs) to estimate the marginal distribution of node variables
for any factor graph. Specifically, we will use BP to approximately compute 
marginals of the distribution (\ref{eq:GraphicalModel}). 

We will quickly recall BP for our specific setup. We refer reader to
see \cite{Pearl, WJ} for further details on the algorithm. BP is a 
message passing algorithm in which at each iteration messages are
sent from variable nodes to neighboring clause nodes and vice versa. The
messages at iteration $t+1$ are functions of messages received at iteration
$t$. To describe the message update equations, we need some notation. Let
$\da$ denote the set of all variables that belong to clause $a$.
If variable $x_i$ is involved in clause $a$ as literal $z$ 
(either $z=x_i$ or $z=\bar{x}_i$), then define  $\dpi(a)$ as the set 
of all clauses (minus $a$) in which $x_i$ appears as $z$. Similarly, 
$\dmi(a)$ denotes the set of all clauses in which $x_i$ appears as $\bar{z}$.
Let $\{h^{(t)}_{i\to a}\}$, $\{u^{(t)}_{a\to i}\}$ denote the messages 
(ideally they are half log-likelihood ratios) that are passed along the ddirected edges 
$i\to a$ and $a\to i$
respectively at time $t$, then  the precise update equations are
\begin{eqnarray}
h_{i\to a}^{(t+1)} = \sum_{b\in\dpi(a)}u^{(t)}_{b\to i}
-\sum_{b\in\dmi(a)}u^{(t)}_{b\to i}\,
, \;\;\;\;\;\;
u^{(t)}_{a\to i}= f(\{h^{(t)}_{j\to a};j\in\da\backslash i\})\, ,
\label{eq:BPUpdate}
\end{eqnarray}
where the function $f(\, \cdot\,)$ has been defined in 
Eq.~(\ref{eq:FDefinition}). We shall assume \footnote{In fact an arbitrary
initial condition and a smaller number of iterations wouldn't change
our main results.} that the update equations are initialized by $h^{(0)}_{i\to a} =0$
and algorithm stops at iteration $t_{\rm max}$ which is equal to the diameter of 
$G_F$. Let $(h_{i\to a}, u_{a\to i})$ be messages passed in the last iteration of BP. 
Using these messages, an estimate of the probability that 
a clause is satisfied can
be obtained as follows. Let $E_a(\ux_{\da})$ be the indicator function for
the $a$-th clause not being satisfied. As mentioned above, 
$h_{i\to a}$ is thought of as half log-likelihood ratio for $i$ satisfying 
$a$ and $i$ not satisfying $a$, in the absence of clause $a$ itself. 
A little algebra then shows that the BP estimate for the expectation
of $E_a(\ux_{\da})$ is 
\begin{eqnarray}
\<E_a(\ux_{\da})\>_{\sBP} = \frac{\sum_{\ux_{\da}}
E_a(\ux_{\da})\; \exp\{-\beta E_a(\ux_{\da})+  h_{i\to a}\sigma_{ai}(x_i)\}}
{\sum_{\ux_{\da}} \exp\{-\beta E_a(\ux_{\da})+ h_{i\to a}\sigma_{ai}(x_i)\}}
\, ,
\end{eqnarray}
where $\sigma_{ai}(x) = +1$ if setting $x_i=x$ satisfies clause $a$,
and $=-1$ otherwise. We further introduce the number of clauses violated by
$\ux$, $E(\ux) = \sum_a E_a(\ux_{\da})$, and its BP estimate 
$\<E(\ux)\>_{\sBP} = \sum_a \<E_a(\ux)\>_{\sBP}$.  

Given $\beta>0$, we let  $\beta_i = i\beta/N^{2}$, for $i=0,\dots,n\equiv N^{2}$. 
Then,
\begin{eqnarray}
\log Z(\beta,F) & = & \log Z(0,F) + 
\sum_{i=0}^{n-1}\log \frac{Z(\beta_{i+1},F)}{Z(\beta_{i},F)} = 
 N\log 2+\sum_{i=0}^{n-1}\log \<e^{-\Delta E(\ux)}\>_i\, ,
\label{eq:Telescopic}
\end{eqnarray} 
where $\Delta \equiv \beta_{i+1}-\beta_i$, and 
$\<\,\cdot\,\>_i$ is a shorthand for expectation with respect
to the measure $\mu_{\beta_i,F}(\,\cdot\,)$. The above 
expression is difficult to evaluate. However, due to $\Delta$
being small the $\<-\Delta E(\ux)\>$ is a good estimate of 
$\log \<e^{-\Delta E(\ux)}\>_i$. Hence, define the algorithm  estimate as
\begin{eqnarray}
\Phi(\beta,F) = N\log 2 - \sum_{i=1}^{n-1}\Delta\, \<E(\ux)\>_{\sBP,i}\, ,
\label{eq:AlgorithmOutput}
\end{eqnarray}
where the subscript in $\<\,\cdot\,\>_{\sBP,i}$ emphasizes that the BP 
computation must be performed at inverse temperature $\beta_i$.

\subsection{Key Lemmas}
\label{ssec2}

Before presenting useful Lemmas, let us mention a few facts. 
Given factor graph $G_F$
and variable node $i$, $1\leq i \leq N$, let $\Ball_i(r)$ denote subgraph
induced by the set of all variable that are within shortest path distance $r$ 
of node $i$ (distance between two variables sharing a clause is unit). 
Analogously, for a clause node $a$,  $\Ball_a(r)$ is the union
of $\Ball_i(r)$ with $i$ running over the variables involved in $a$. Let
$A$ be subset of variable nodes. Then,  let $\ux_A$ denote an 
assignment to the corresponding variables. Given two such subsets 
$A,B\subseteq [N]$ and assignments $\ux_A$, $\ux_B$, let 
$\mu_{A|B}(\ux_A|\ux_B)$ be the conditional probability under 
the distribution (\ref{eq:GraphicalModel}) of the variables in $A$, 
given assignment $\ux_B$ on $B$. The following is a well-known
result about BP algorithm (see \cite{TJ99}). 
\begin{lemma}\label{lemma:BoundBP}
Given a clause $a$ and $r$, let   $\Ball_a(r+1)$ be a tree.
Let $U = \Ball_a(r)$ and $V=[N]\backslash U$. Then 
\begin{eqnarray}
\left|\<E_a(\ux_{\da})\>-\<E_a(\ux_{\da})\>_{\sBP}\right|
\le \sup_{\uy,\uz}\l|\mu_{\da|V}(\,\cdot\,|\uy_{V})-
\mu_{\da|V}(\,\cdot\,|\uz_{V})\r|_{\sTV}\, ,\\
0\le\<E_a(\ux_{\da})\>,\,\<E_a(\ux_{\da})\>_{\sBP}\le
\max_{\uz_{V}}\left\{\sum_{\ux_{\da}}E_a(\ux)\mu_{\da|V}(\ux_{\da}|\uz_V)
\right\}
\end{eqnarray}
\end{lemma}

Next, we present a known result about locally tree-like structure of random
$k$-SAT formula (an analogous result concerns the local structure of
sparse random graphs).  
\begin{lemma}\label{lemma:LocalTree}
Consider $k\ge 2$, $\alpha\in [0,\infty)$ and a random $k$-SAT formula
$F$ with clause density $\alpha$. For $r\ge 0$, let $\Ball_i(r)$ be the ball
of radius $r$ centered at a uniformly random variable node $i$. Let 
$S(r)$ be an $r$-generation tree with distribution same as  $\T_*(r)$
(with the same values of $k$ and $\alpha$). Then, there exists $A, \rho$ (dependent
on $\alpha, k$) such that
\begin{eqnarray}
\l|\,\prob\{\Ball_i(r)\in\,\cdot\, \}-\prob\{S(r)\in\,\cdot\, \}\,
\r|_{\sTV}\le \frac{A\,e^{\rho r}}{N}\, .
\end{eqnarray}
\end{lemma}
\begin{figure}
\center{\includegraphics[width=6.cm]{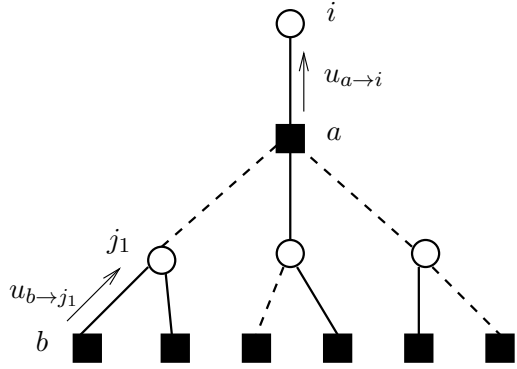}}
\put(-182,5){$b$}
\put(-192,25){$u_{b\to j_1}$}
\put(-155,44){$j_1$}
\put(-72,85){$a$}
\put(-73,107){$u_{a\to i}$}
\put(-72,130){$i$}
\caption{Pictorial representation of the recursion (\ref{eq:RecursionU})
on the factor graph $G_F$: filled squares represent function nodes and
empty circles variable nodes. Dashed edges correspond to negations.}
\label{fig:TwoGen}
\end{figure}

\begin{lemma}\label{lemma:Key}
Let $\alpha_*(k)$ be  the smallest positive root of the equation 
$\kappa(\alpha) = 1$, with $\kappa(\alpha)$ defined as 
in Eq.~(\ref{eq:ContractionRate}).
Then  $\alpha_*(k)\le\alpha_{\rm u}(k)$.
\end{lemma}
\prooft
Given an $r$-generations tree formula $F$, consider an edge 
$i\to a$ directed toward the root
and the subtree rooted at $i$ and \emph{not} containing $a$. 
Denote by $\mu_{i\to a}( \, \cdot\,)$ the marginal distribution of
$x_i$ with respect to the model associated to this subtree, and 
let $h_{i\to a}\in[-\infty,\infty]$ be the corresponding log-likelihood ratio
\begin{eqnarray}
h_{i\to a} \equiv \frac{1}{2}\log \left\{
\frac{\mu_{i\to a}(\mbox{$x_i$ satisfies $a$})}
{\mu_{i\to a}(\mbox{$x_i$ doesn't satisfy $a$})}\right\}\, .\label{eq:DefLLR}
\end{eqnarray}
Analogously, given an edge $a\to i$, we consider the subtree rooted at $i$ and 
 containing  only $a$ among the clauses involving $i$. We denote 
by $\mu_{a\to i}( \, \cdot\,)$ the corresponding marginal distribution at 
$i$, and let
\begin{eqnarray}
u_{a\to i} \equiv \frac{1}{2}\log \left\{
\frac{\mu_{a\to i}(\mbox{$x_i$ satisfies $a$})}
{\mu_{a\to i}(\mbox{$x_i$ doesn't satisfy $a$})}\right\}\, .
\end{eqnarray}
It is easy to show that these log-likelihoods satisfy the 
recursions\footnote{The reader will notice that these coincide with 
the BP update equations, cf. Eq.~(\ref{eq:BPUpdate}), which are known to 
be exact on trees.}
\begin{eqnarray}
h_{j\to a} = \sum_{b\in\dpj(a)}u_{b\to j}
-\sum_{b\in\dmj(a)}u_{b\to j}\,
, \;\;\;\;\;\;\;\;\;\;
u_{a\to i}= f(\{h_{j\to a};j\in\da\backslash i\})\, ,
\end{eqnarray}
with the function $f(\,\cdot\,)$ being defined as in 
Eq.~(\ref{eq:FDefinition}).
For the calculations below, it is convernient 
to eliminate the $h_{i\to a}$ variables, to get
\begin{eqnarray}
u_{a\to i}= f\left( \sum_{b\in\dpj_1(a)}u_{b\to j_1}\!
-\!\!\!\sum_{b\in\dmj_1(a)}u_{b\to j_1};\;\; \dots\;\;;
\sum_{b\in\dpj_{k-1}(a)}u_{b\to j_{k-1}}\!
-\!\!\!\sum_{b\in\dmj_{k-1}(a)}u_{b\to j_{k-1}}\right)\, ,\label{eq:RecursionU}
\end{eqnarray}
where we denoted by $j_1,\dots, j_{k-1}$ the indices of variables involved in
clause $a$ (other than $i$).
A pictorial representation of this recursion is provided in 
Fig.~\ref{fig:TwoGen}.

Notice that the above recursions hold irrespective whether
one considers the unconditional measure $\mu(\,\cdot\,)$, or 
the conditional one $\mu(\, \cdot\,|\ux_r)$. What changes
in the two cases are the initial condition for the recursion,
i.e. the value of $h_{i\to a}$ associated with the variables 
$i$ at the $r$-th generation. For the unconditioned measure
(`free boudary condition'), the appropriate initialization is $h_{i\to a}=0$.
If one conditions to $\ux_r$, $h_{i\to a}= +\infty$, or
$=-\infty$ depending (respectively) whether $x_i$ satisfy clause
$a$ or not.

In the rest of the proof, we shall think always to the conditioned 
measure $\mu(\, \cdot\,|\ux_r)$. As a consequence, the log-likelihoods
are, implicity, functions of $\ux_r$: 
 $u_{a\to i} = u_{a\to i}(\ux_r)$ 
(indeed of the restriction of $\ux_r$  to the subtree rooted at $i$,
and only containing $a$). 
We then let
\begin{eqnarray}
\ou_{a\to i} = \max_{\ux_r}\; u_{a\to i}(\ux_r)\, ,\;\;\;\;\;
\uu_{a\to i} = \min_{\ux_r}\; u_{a\to i}(\ux_r)\, .
\end{eqnarray}
In the case $\beta = \infty$, the maximum (minimum) is taken over 
all boundary conditions $\ux_r$, such that the sub-formula rooted
at $i$ admits at least one solutions, under the condition $\ux_r$
(there is always at least one such boundaries).
We further let $\Delta_{a\to i} = \ou_{a\to i}-\uu_{a\to i}\ge 0$.

Consider a random tree distributed as $\T_*(r)$, conditioned to the root 
having degree $1$, i.e. to the root variable being involved in a unique 
clause, to
be denoted by $a$. Let $\Delta^{(r)} = \Delta_{a\to i}$ be the corresponding
log-likelihoods interval. We will show that 
$\E\tanh\Delta^{(r)}\le  e^{-\gamma r}$ for some positive constant
$\gamma$. Before proving this claim, let 
us show that it indeed implies the thesis.
Denoting by $\partial_+0$ the set of clauses in which the root is involved
as the direct literal, and by $\partial_-0$ the set in which it is
involved as negated, we have
\begin{eqnarray}
\l|\mu_{0|r}(\,\cdot\,|\ux_r)-\mu_{0|r}(\,\cdot\,|\uz_r)\r|_{\sTV} &=& 
\frac{1}{2}\big|\tanh h_0(\ux_r)-\tanh h_0(\uz_r)\big| \, ,\\
h_0(\ux_r)  &\equiv &\sum_{a\in\partial_+0} u_{a\to 0}(\ux_r)-
\sum_{a\in\partial_-0} u_{a\to 0}(\ux_r)\, .
\end{eqnarray}
Since $x\mapsto \tanh(x)$ is monotonically increasing in $x$, we have
\begin{eqnarray}
&&\l|\mu_{0|r}(\,\cdot\,|\ux_r)-\mu_{0|r}(\,\cdot\,|\uz_r)\r|_{\sTV} \le 
\frac{1}{2}\big\{\tanh \oh_0-\tanh \uh_0\big\} \, ,\\
&&\oh_0  \equiv \sum_{a\in\partial_+0} \ou_{a\to 0}-
\sum_{a\in\partial_-0} \uu_{a\to 0}\, ,\;\;\;\;\;\;\;\;
\uh_0  \equiv \sum_{a\in\partial_+0} \uu_{a\to 0}-
\sum_{a\in\partial_-0} \ou_{a\to 0}\, .\label{eq:h0}
\end{eqnarray}
Using the elementary properties $\tanh x-\tanh y\le 2\tanh(x-y)$
for any $x\ge y$, and $\tanh(x+y)\le \tanh x+\tanh y$ for $x,y\ge 0$,
we get
\begin{eqnarray}
\l|\mu_{0|r}(\,\cdot\,|\ux_r)-\mu_{0|r}(\,\cdot\,|\uz_r)\r|_{\sTV} \le  
\tanh\left\{\sum_{a\in\partial 0}\Delta_{a\to 0}\right\} 
 \le \sum_{a\in\partial 0} \tanh \Delta_{a\to 0}\, .
\end{eqnarray}
We can take the maximum over boundary condition and
the expectation with respect to the tree ensemble. Recalling
that $|\partial 0|$ is a Poisson random variable of mean $k\alpha$, we get
\begin{eqnarray}
\E\max_{\ux,\uz}  \l|\mu_{0|r}(\,\cdot\,|\ux_r)-\mu_{0|r}
(\,\cdot\,|\uz_r)\r|_{\sTV} \le  k\alpha\, \E
\tanh\Delta^{(r)} \, ,
\end{eqnarray}
which implies the thesis upon taking $A=k\alpha $.

We are now left with the task of proving $\E\tanh\Delta^{(r)}\le 
e^{-\gamma r}$. It is easy to realize that $f(x_1,\dots, x_{k-1})$
is monotonically decreasing in each of its arguments.
Therefore Eq.~(\ref{eq:RecursionU}) yields the following recursion 
for upper/lower bounds
\begin{eqnarray}
\ou_{a\to i}=f\left( \sum_{b\in\dpj_{1}(a)}\uu_{b\to j_1}\!
-\!\!\!\sum_{b\in\dmj_1(a)}\ou_{b\to j_1};\;\; \dots\;\;;
\sum_{b\in\dpj_{k-1}(a)}\uu_{b\to j_{k-1}}\!
-\!\!\!\sum_{b\in\dmj_{k-1}(a)}\ou_{b\to j_{k-1}}\right)\, ,
\end{eqnarray}
together with the equation obtained by interchanging $\uu_{\cdots}$
and $\ou_{\cdots}$. By taking the difference of these two equations, we
get
\begin{eqnarray}
\Delta_{a\to i} = f(\uh_{1};\;\dots\; ;\uh_{k-1})-
f(\oh_{1};\;\dots\; ;\oh_{k-1})\, ,
\end{eqnarray}
where we defined $\uh_{i} = \sum_{b\in\dpj_{i}(a)}\uu_{b\to j_i}\!
-\!\!\!\sum_{b\in\dmj_i(a)}\ou_{b\to j_i}$
and $\oh_{i} = \sum_{b\in\dpj_{i}(a)}\ou_{b\to j_i}\!
-\!\!\!\sum_{b\in\dmj_i(a)}\uu_{b\to j_i}$
(obviously $\uh_i\ge \oh_i$).

Suppose now $n$ out of the $k-1$ variables $x_{j_1},\dots, x_{j_{k-1}}$
are pure literals, let's say variables  $x_{j_1},\dots,x_{j_n}$. 
This means that $\dmj_1(a), \dots \dmj_n(a) = \emptyset$, and therefore,
since the loglikelihoods $u_{b\to j}$ are non-negative (because $f$ is 
non-negative), $\uh_{1},\dots \uh_n\ge 0$.
It is an easy exercise of analysis to show that, if $x_1,\dots,x_n\ge 0$,
\begin{eqnarray}
0\le-\frac{\partial f}{\partial x_i}(x_1,\dots,x_{k-1})\le \frac{1}{2^n}\, .
\end{eqnarray}
Therefore, by the Mean Value Theorem
\begin{eqnarray}
\Delta_{a\to i} \le \frac{1}{2^n}\sum_{l=1}^{k-1} (\oh_l-\uh_l)
=\frac{1}{2^n}\sum_{l=1}^{k-1} \sum_{b\in\partial j_l}
\Delta_{b\to j_l}
\, ,
\end{eqnarray}
Next we take the hyperbolic tangent of both sides, and 
use again $\tanh(x+y)\le \tanh x+\tanh y$, for $x,y\ge 0$
to get 
\begin{eqnarray}
\tanh\Delta_{a\to i} \le \frac{1}{2^n}\sum_{l=1}^{k-1} \sum_{b\in\partial j_l}
\tanh \Delta_{b\to j_l}
\, .
\end{eqnarray}
Finally we take expectation of this inequality. In order to do this,
we recall that $n$ is just the number of pure literals
among $x_{j_1},\dots x_{j_{k-1}}$. In our notations this can be written
as $n = \sum_{l=1}^{k-1}\ind(|\dmj_l(a)| =0)$. We further assume that 
$i$ is the root of a tree from $\T_*(r+1)$, $r\ge 0$ and therefore 
$\Delta_{a\to i}$ is distributed as $\Delta^{(r)}$. Furthermore
the differences $\Delta_{b\to j_{l}}$ will be distributed as 
$\Delta^{(r+1)}$. We thus obtain
\begin{eqnarray}
\E\tanh\Delta^{(r+1)} &\le &\E\left\{
\prod_{l=1}^{k-1}\frac{1}{2^{\ind(|\dmj_l(a)| =0)}} 
\sum_{l=1}^{k-1}
\sum_{b\in\partial j_l(a)}
\tanh \Delta^{(r)}\right\} =\\
&=& (k-1)\,\E\left\{\frac{1}{2^{\ind(|\dmj| =0)}}
|\partial j|\right\}\, 
\left\{\E\,2^{-\ind(|\dmj| =0)}\right\}^{k-2}
\, \E\tanh\Delta^{(r)} 
\, .
\end{eqnarray}
The expectations over $|\dpj|$, $|\dmj|$ are easily evaluated by recalling that
these are inpependent Poisson random variables of mean $k\alpha/2$.
One finally obtains $\E\tanh\Delta^{(r+1)} \le\kappa(\alpha) 
\E\tanh\Delta^{(r)}$. The thesis follows (with $\gamma = -\log\kappa(\alpha)$)
by noticing that $\E\tanh \Delta^{(0)}\le 1$, and recalling that 
$\kappa(\alpha)<1$ for $\alpha<\alpha_*(k)$.
\endproof

Next, we state result about the error in expectation w.r.t. to BP estimate in
a clauses being satisfied or not.  To obtain bound in the error of BP estimate of $\<E_a(\ux)\>$, 
we need to study the error in estimation of the joint distribution 
of $k$ variables in a clause. For this, we first choose a clause at 
random and treat all of its $k$ variables as root of $k$ independent rooted random 
trees (of suitable depth $r$) as before. Note that, this asymptotically
does not bias the distribution of the original random formula as
this process tilt the original distribution by at most $O(1/N)$.

To this end,  let $\ux_r$ be an 
assigment for the $r$-th generation variables. We shall denote by 
$\<\,\cdot\,\>^{(r)}$ the expectation with respect to the graphical model 
(\ref{eq:GraphicalModel}) associated to a formula constructed as follows.
First we generate a uniformly random clause over variables $x_1,\dots, x_k$. 
Then we sample $k$ independent trees according to $\T_*(r)$ and root them
at $x_1,\dots, x_k$.
We let $\<\,\cdot\,\>_{\ux_r}^{(r)}$ be 
the corresponding conditional expectation,
given the assignment to the $r$-th generation.
\begin{lemma}\label{lemma:EnergyTree}
Let $k\ge 2$, $\alpha<\alpha_*(k)$ and $\beta\in[0,\infty]$.
Then there exist two positive constants $A$, $\gamma$, such that
\begin{eqnarray}
\E\max_{\ux_r,\uz_r}
\left|\<E_a(\ux)\>_{\ux_r}^{(r)}-\<E_a(\ux)\>_{\uz_r}^{(r)}\right|\le 
A\, e^{-\gamma r}\, .
\end{eqnarray}
\end{lemma}
\prooft
Denote by $\ux_{\da} = \{x_1,\dots ,x_{k}\}$ the zeroth generation 
variables, by $T_1,\dots, T_k$ the tree factor graphs drawn from $\T_*(r)$ and rooted,
respectively, at variable nodes $1,\dots,k$.
We then denote by  $\mu_i(x_i|\ux_r)$, $i\in\{1,\dots,k\}$ the conditional 
distribution for variable $x_i$ with respect to the model associated with the 
tree $T_i$. We also let 
$h_i(\ux_r)$ be the associated log-likelihoods 
(defined analogously to Eq.~(\ref{eq:DefLLR})), and
$\uh_i = \max_{\ux_r}h_i(\ux_r)$ ($\oh_i = \min_{\ux_r}h_i(\ux_r)$)
be their maximum (minimum) values with respect to the boundary condition.

It is not hard  to show that
$\<E_a(\ux)\>_{\ux_r}^{(r)} = g(h_1(\ux_r),\dots,h_k(\ux_r))$
where the function $g:\R^k\to\R$ is defined as follows
\begin{eqnarray}
g(x_1,\dots,x_k) \equiv \frac{e^{-\beta}\prod_{i=1}^k\frac{1}{2}(1-\tanh x_i)}
{1-(1-e^{-\beta})\prod_{i=1}^k\frac{1}{2}(1-\tanh x_i)}\, .
\label{eq:Gdef}
\end{eqnarray}
Since $g(x_1,\dots,x_k)$ is monotonically decreasing in each of its arguments,
we have
\begin{eqnarray}
\E\max_{\ux_r,\uz_r}
\left|\<E_a(\ux)\>_{\ux_r}^{(r)}-\<E_a(\ux)\>_{\uz_r}^{(r)}\right|
\le \E\left\{g(\uh_1,\dots,\uh_k)- g(\oh_1,\dots,\oh_k)\right\}\, ,
\label{eq:GDiff}
\end{eqnarray}
where the couples  $(\uh_1,\oh_1),\dots,(\uh_k,\oh_k)$ are i.i.d.'s
and distributed as $(\uh_0,\oh_0)$ in the proof of Lemma  \ref{lemma:Key},
cf. Eq.~(\ref{eq:h0}). In particular, proceeding as in that proof,
we deduce that $\E\tanh(\oh_i-\uh_i)\le A\, e^{-\gamma r}$.
We are left with the task of proving that this implies an analogous bound
on the right hand side of Eq.~(\ref{eq:GDiff}).

To this end, we first consider a single variable function $\gt:\R\to \R$
with $0\le \gt(x)\le 1$ and $-1\le\gt'(x)\le 0$. Then
\begin{eqnarray}
\E\{\gt(\uh_1)-\gt(\oh_1)\} & \le & \prob\{\oh_1-\uh_1\ge \Delta\}+
\E\{(\oh_1-\uh_1)\, \ind(\oh_1-\uh_1<\Delta)\}\le\\
&\le &\frac{1}{\tanh\Delta}\E\tanh(\oh_1-\uh_1)+\frac{\Delta}{\tanh\Delta}
\E\{\tanh(\oh_1-\uh_1)\, \ind(\oh_1-\uh_1<\Delta)\}\le \nonumber\\
&\le&\frac{1+\Delta}{\tanh\Delta}\E\tanh(\oh_1-\uh_1)\, .\nonumber
\end{eqnarray}
The proof is completed by writing 
$\E\{g(\uh_1,\dots,\uh_k)- g(\oh_1,\dots,\oh_k)\} =
\sum_{i=0}^k \E\{\gt_i(\uh_i)-\gt_i(\oh_i)\}$ where
$\gt_i(x) \equiv g(\oh_1\dots \oh_{i-1},x,\uh_{i+1},\dots,\uh_k)$ and 
noticing that $-1\le\frac{\partial g}{\partial x_i}\le 0$ (the last statement 
is proved in the appendix)
\endproof
%
\iffalse
Because of Lemma \ref{lemma:BoundBP}, the above two results, namely 
Lemmas \ref{lemma:LocalTree} and  \ref{lemma:EnergyTree}, imply that belief 
propagation provides a good approximation of averages with 
respect to the measure $\mu_{\beta,F}(\,\cdot\,)$.
For the sake of concreteness we limit ourselves to bounding the
expected error on the average cost $\<E(\ux)\>$. Any local marginal can
however be treated along the same lines.
%
\fi

Finally, a result that puts together the above observations to derive the net 
error in BP estimation.
\begin{lemma}\label{lemma:EnergyEstimate}
Let $k\ge 2$, $\alpha<\alpha_*(k)$ and $\beta\in[0,\infty]$. 
Then there exists two  positive constants $C$ and $\delta<1$
such that for any $N$, 
\begin{eqnarray}
\E\left|\<E(\ux)\>-\<E(\ux)\>_{\sBP}\right|\le CN^{\delta}\, .
\label{eq:BoundTotalEnergy}
\end{eqnarray}
\end{lemma}
\prooft
By linearity of expectation and using Lemma \ref{lemma:BoundBP},
we get
\begin{eqnarray}
\E\left|\<E(\ux)\>-\<E(\ux)\>_{\sBP}\right| \le 
M\E\left|\<E_a(\ux)\>-\<E_a(\ux)\>_{\sBP}\right|\le 
M\E\left\{\max_{\ux,\uz}\left|\<E_a(\ux)\>^{(r)}_{\ux_r}-
\<E_a(\ux)\>^{(r)}_{\uz_r}\right|\right\}\, .
\end{eqnarray}
We would like to apply Lemma \ref{lemma:EnergyTree}, but the expectation
in the last expression is taken with respect to the formula $F$ drawn from the
random $k$-SAT ensemble, instead of the tree model $\hT_*(r)$. However,
the quantity in curly brackets depends only of the radius $r$ neighborhood
$\Ball_a(r)$ of vertex $a$ in $G_F$. Furthermore is non negative and 
upper bounded by 1. We can therefore apply Lemma
\ref{lemma:LocalTree} and \ref{lemma:EnergyTree} to upper bound the last 
expression by (here 
$\E_{\hT}$ denotes expectation with respect to the tree ensemble):
\begin{align}
M\l|\,\prob\{\Ball_i(r)\in\,\cdot\, \}-\prob\{S(r)\in\,\cdot\, \}\,
\r|_{\sTV}+ M\E_{\hT}\left\{\max_{\ux,\uz}\left|\<E_a(\ux)\>^{(r)}_{\ux_r}-
\<E_a(\ux)\>^{(r)}_{\uz_r}\right|\right\}&\le\\
\le A\alpha\,& e^{\rho r}+NA'\alpha \, e^{-\gamma r}\nonumber
\end{align}
The proof is completed by setting $r = \frac{1}{\rho+\gamma}\log N$,
which yields Eq.~(\ref{eq:BoundTotalEnergy}) with $\delta = 
\frac{\rho}{\gamma+\rho}$.
\endproof
%.

\section{Proofs of Theorems}
\label{sec3}

\subsection{Proof of Theorem \ref{thm:KSAT}}
Clearly, the running time of algorithm described in Section \ref{sec2} is
$O(N^4)$ as total number of BP runs are $O(N^2)$ and each BP 
run takes $O(N)$ iterations or $O(N^2)$ serial operations. Now,
we'll prove Eq.~(\ref{eq:Guarantee}).

Using te existing lower bounds on $\alpha_{\rm c}(k,N)$ (see \cite{AP04}
and references therein), it is not hard to show that 
$\alpha_*(k)\le\alpha_{\rm c}(k,N)(1-\eta)$ for some $\eta>0$
all $k\ge 2$ and $N$ large enough.
By definition, for $\alpha<\alpha_{\rm c}(k,N)(1-\eta)$,
$\beta\in[0,\infty]$ there exists a constant $C(\alpha)>0$ such that
$\log Z(\beta,F)\ge C(\alpha)N\log 2$ whp.  This follows from the following
two facts for appropriate $C(\alpha)$: (1) at least $C(\alpha)N$ variables  
do not appear in any clause whp and (2) at least one solution is satisfying
assignment whp as $\alpha < \alpha_{\rm c}(k,N)(1-\eta)$.  Thus, there are
at least $2^{C(\alpha) N}$ satisfying assignment, whence 
$\Z(\beta, F) \geq 2^{C(\alpha) N}$.
Given this, it is sufficient to show that 
$\left|\log Z(\beta,F)-\Phi(\beta,F)\right|\le N\ve$ w.h.p.
for any $\ve>0$ and $N$ large enough. 

Now, Eqs.~(\ref{eq:Telescopic}) and (\ref{eq:AlgorithmOutput}) imply that
\begin{eqnarray}
\left|\log Z(\beta,F)-\Phi(\beta,F)\right| &\le & 
\sum_{i=0}^{n-1}\left|\log \<e^{-\Delta E(\ux)}\>_i+
\Delta\, \<E(\ux)\>_{\sBP,i}\right|\nonumber\\
&\le & 
\sum_{i=0}^{n-1}\left|\log \<e^{-\Delta E(\ux)}\>_i+
\Delta\, \<E(\ux)\>_{i}\right|+\sum_{i=0}^{n-1}
\Delta\left|\<E(\ux)\>_{i}-\<E(\ux)\>_{\sBP,i}\right|\, .
\label{eq:BoundDiff}
\end{eqnarray}
Consider the first term in (\ref{eq:BoundDiff}): for any non-negative
random variable $X$, 
$ \log \< e^{-X} \> \leq \< e^{-X} \> - 1  \leq$   $\<1 - X + X^2 \> - 1 \leq - \< X \> + \<X^2 \>$.
As a consequence, we obtain 
\begin{eqnarray}
\sum_{i=0}^{n-1}\left|\log \<e^{-\Delta E(\ux)}\>_i+
\Delta\, \<E(\ux)\>_{i}\right| ~\leq~\sum_{i=0}^{n-1}\Delta^2\<E(\ux)^2\>_i\le \beta\Delta\sup_i\<E(\ux)^2\>_i\le 
N^{2\delta'}\alpha^2\, ,
\end{eqnarray}
where we used $\beta\le N^{\delta'}$, $\Delta = \beta/N^2\le N^{\delta'-2}$
and $0\le E(\ux)\le N\alpha$. If we choose $\delta'<1/2$, this
contribution is smaller than $N\ve/2$ for all $N$ large enough.

Now, the second term in Eq.~(\ref{eq:BoundDiff}): the bound 
(\ref{eq:BoundTotalEnergy}) holds for any $\beta$ in the compact region 
$[0,\infty]$. Furhter, the left hand side is uniformly bounded (in terms of $N$) and 
continuous in $\beta$. Hence, there exists a $C$ so that the bound (\ref{eq:BoundTotalEnergy})
holds uniformly for $\beta\in[0,\infty]$. This will imply that
\begin{eqnarray}
\sum_{i=0}^{n-1}
\Delta\,\E\left|\<E(\ux)\>_{i}-\<E(\ux)\>_{\sBP,i}\right|
&\le & \beta C N^{\delta} \le C N^{\delta+\delta'}
\end{eqnarray}
Choosing $\delta'\in (0,1-\delta)$ and Markov inequality will imply
that the second term is also bounded above by $N\ve/2$ whp. This
completes the proof of Theorem \ref{thm:KSAT}.
\endproof
%
%************************************************************************
%
\subsection{Proofs of Theorems \ref{thm:UniquenessTrees}, \ref{thm:LimitPartFun},
and \ref{thm:AlmostSAT}}
\label{sec:Sketch}

Due to shortage of space, they are moved to Appendix \ref{ap1}.

\section{Discussion and Future Work}
\label{sec4}

We presented a novel deterministic algorithm for approximately counting 
good truth assignments
of random $k$-SAT formula with high probability. The algorithm is built upon the well-known
Belief Propagation heuristic and an interpolation method for the log-partition function. In
the process of establishing the correctness of the algorithm, we obtained the threshold
for uniqueness of Gibbs distribution for random $k$-SAT formula as $2k^{-1}\log k (1+o_k(1))$. 
This result if of interest in its own right. 

We believe that our result can be extended to a reasonable class of non-random $k$-SAT
formula. We also believe that the approximation guarantees of Theorem \ref{thm:KSAT} 
should hold for any $\beta \in [0,\infty]$. 

%
%************************************************************************
%

%
%************************************************************************
%
\appendix
\section{Proof Sketches: Theorems \ref{thm:UniquenessTrees}, \ref{thm:LimitPartFun} and
\ref{thm:AlmostSAT}}
\label{ap1}

Due to space limitations, we only provide sketch of  proofs  for
Theorems  \ref{thm:UniquenessTrees}, \ref{thm:LimitPartFun} and
\ref{thm:AlmostSAT}.

\vspace{.1in}
\noindent{\bf Proof sketch for Theorem \ref{thm:UniquenessTrees}.} 
By using the definition
$\kappa(\alpha_*) = 1$ (with $\kappa(\alpha)$ being defined as in 
Eq.~(\ref{eq:ContractionRate})), it is easy to show that 
$\alpha_*(k) = 2k^{-1}\log k\{1+O(\log\log k/\log k)\}$. To complete
the proof, we need a (asymptotically in $k$) matching upper bound. 
In order to obtain such an upper bound, we consider the case $\beta =\infty$, i.e. only
satisfying assignments have positive weight.  Consider  a tree formula
which is distributed as $\T_*(r)$. Let $P_r$ be the probability 
that there exists two boundary conditions $\ux^{(0)}_r$, $\ux^{(1)}_r$,
such that the root takes values, respectively, $0$ or $1$ in all the 
satisfying assignments with the respective boundary conditions. 
Clearly for the Gibbs measure to be unique (or have correlation decay) 
in the sense of  Definition \ref{def:Uniqueness}
(but also in the weaker sense correspondint to the threshold
$\alpha'_{\rm u}(k)$), it must be that $P_r\to 0$ 
as $r\to\infty$. Hence, if we establish that for 
$ \alpha > 2k^{-1}\log k\{1+O(\log\log k/\log k)\}$,
there exists such boundary conditions with positive probability, then the
proof will be complete. Next, we do that. 

For this, consider a tree from $\T_*(r)$ with the root having degree
$1$. Given 
such a tree, let $\rho_r$ be the probability that there 
exists a boundary condition $\ux_r$, such that the root variable is 
the only variable that satisfies the only clause in which it belongs (recall
that the root variable has degree $1$) for all possible satisfying assignments
with the given boundary condition. If $P_r\to 0$, then  $\rho_r\to 0$.
To prove this claim, assume by contraddiction that $\rho_r$ remains 
bounded away from zero (say $\rho_r\ge \underline{\rho}>0$) and consider
an tree from  $\T_*(r)$ (without conditioning). With finite probability
the root belongs to two clauses in which it appears, respectively,
directed and negated. With probability at least $\underline{\rho}^2>0$, for 
each of the corresponding subtrees there exists a boundary condition 
that fixes the root variable to be (respectively) directed or negated.
By extending arbitrarily this boundary conditions to the full tree,
we obtain the desired $\ux_r^{(1)}$, $\ux_r^{(0)}$. 

It turns out that $\rho_r$ can be determined recursively. Set $\rho_0=1$
and $\rho_{r+1} = \{1-\exp(-k\alpha\rho_r/2)\}^{k-1}$. Recursively, $\rho_r \to 0$
as $r\to\infty$ only if $\alpha < \alpha^*(k)$, where $\alpha^*(k)$ for the
above recursion (with little bit of algebra) evaluates to 
$\alpha^*(k) = 2k^{-1}\log k\{1+O(\log\log k/\log k)\}$. This completes
the proof sketch of Theorem \ref{thm:UniquenessTrees}. 

\vspace{.1in}
\noindent{\bf Proof sketch for Theorem \ref{thm:LimitPartFun}.} First notice
that, if $F$ and $F'$ differ in a single clause, then $|\log Z(\beta,F)-\log Z(\beta,F')|\le 2\beta$. 
Hence, by application of Azuma-Hoeffding's inequality, it follows that 
$|\log Z-\E\log Z|\le N\delta$ with probability at least $1-e^{-NC_\beta \delta^2}$, 
for some $C_\beta >0$ for any $\beta \in [0,\infty)$.  Given this, to obtain the almost
sure convergence as in (\ref{eq:AlmostSure}), it is sufficient to prove that 
$\lim_{N\to\infty}N^{-1}\E\,\Phi(\beta,F) = \phi(\beta)$,
in light of Theorem \ref{thm:KSAT} and Borel-Cantelli's Lemma.

To do so, first we need to establish that 
\begin{eqnarray}
\lim_{N\to\infty}\frac{1}{N}\E\<E(\ux)\>_{\sBP,\beta} = \alpha
\E g(h_1,\dots,h_k)  \, ,\label{eq:LimitEnergy}
\end{eqnarray}
where $g$ is defined as in Eq.~(\ref{eq:Gdef}); the random variables $h_1,\dots,h_k$
are i.i.d. with distribution $\nu^*$ that is fixed point of operator $S$ as defined in
the statement of Theorem \ref{thm:LimitPartFun}. We claimed that the fixed point
is unique for $S$. To justify this claim, first note that the  
image of $S$ is contained in the space of distributions supported on
$[0,\beta/2]$, call it ${\cal D}_{\beta}$, which is a compact space with respect to
the weak topology. Being continuous on ${\cal D}_{\beta}$, $S$  admits at least one
fixed point in it. Moreover,
the contraction condition implied by the correlation decay (proved as a part of
Theorem \ref{thm:UniquenessTrees}) implies the attractiveness as well as 
the uniqueness of the fixed point of $S$. 

Once we establish existence of the unique fixed point, the (\ref{eq:LimitEnergy})
follows from Lemma \ref{lemma:LocalTree} and correlation decay established in
Theorem \ref{thm:UniquenessTrees}. Now, by integrating Eq.~(\ref{eq:LimitEnergy}) 
over $\beta$ and observing that $\beta_{i+1} -\beta_i = \beta/N^2$ (hence integration
error is negligible at scale $1/N$) one gets
\begin{eqnarray}
\lim_{N\to\infty}N^{-1}\E\,\Phi(\beta,F) = \log 2-\alpha\int_{0}^{\beta}
\E_{\beta'} g(h_1,\dots,h_k)\, {\rm d}\beta'\, ,
\end{eqnarray}
where a subscript has been added in $\E_{\beta'}$ to stress that  the fixed
point distribution has to be taken at inverse temperature $\beta'$.
The proof of Theorem \ref{thm:LimitPartFun} is completed
by showing that the integral on the
right hand side of the last equation is given by $\phi(\beta)$ 
as in Eq.~(\ref{eq:phi}). In fact, by taking the derivative of this expression
wrt $\beta$, one gets a contribution coming from the explicit $\beta$
dependence, which evaluates to $-\alpha\E g(h_1,\dots,h_k)$, and one from
the $\beta$ dependence of the fixed poit distribution, that can be shown 
to vanish.

\vspace{.1in}
\noindent{\bf Proof Sketch of Theorem \ref{thm:AlmostSAT}. }
For the ease of notation,  let $Z(\beta) \equiv \Z(\beta,F)$, $\Xi(\zeta) \equiv\Xi(\zeta,F)$ and
$U(\beta)\equiv\<E(\ux)\>_{\beta,F}$.  Because of Theorem \ref{thm:KSAT}, it is sufficient to prove
that $|\log \Xi(N\epsilon)-\log Z(\beta)|\le N\epsilon^a$ whp. 
This follows from two inequalities.

First inequality. For any $\zeta \geq 0$, 
\begin{eqnarray}
Z(\beta) & = & \sum_{\ux: E(\ux) \geq \zeta} e^{-\beta E(\ux)} + \sum_{\ux: E(\ux) < \zeta} e^{-\beta E(\ux)} 
                \geq  e^{-\beta \zeta} \Xi(\zeta). \label{eq:zz2}
\end{eqnarray}
Second inequality. For any $\zeta \geq 0$ and using the first equality in (\ref{eq:zz2}), we
obtain
\begin{eqnarray}
Z(\beta) & \leq & \sum_{\ux: E(\ux) \geq \zeta} e^{-\beta E(\ux)} + \Xi(\zeta). \nonumber 
\end{eqnarray}
Equivalently, 
$Z(\beta) \mu(E(\ux) < \zeta)  \leq \Xi(\zeta)$.
Now, take $\zeta = 2U(\beta)$ then, we get using Markov's inequality
\begin{eqnarray}
 \mu\{E(\ux) < 2 U(\beta) \} & \geq & 1 - \frac{U(\beta)}{2U(\beta)} ~=~ \frac{1}{2}. \label{eq:zz4}
\end{eqnarray}
>From (\ref{eq:zz2}) and (\ref{eq:zz4}), we obtain
\begin{eqnarray}
 \log Z(\beta) - \log 2 & \leq &  \log \Xi(2U(\beta)) ~\leq~ \log Z(\beta) + 2\beta U(\beta). \label{eq:zz5}
\end{eqnarray}

The next sep consists in controlling $U(\beta)$ at large $\beta$.
Arguing analogously to the proof of Theorem \ref{thm:KSAT}
one can show that there exist constants $C_1$, $C_2$, $C_3$, $a>0$ such 
that, for any $\beta\in [0,\infty]$,
$NC_1 e^{-2\beta}\le U(\beta)\le N C_2e^{-b\beta}+C_3N^{\delta}$
whp. 

Fix $\beta_1$ in such a way that $2C_1e^{-2\beta_1} = \ve$. 
Then $2U(\beta_1)\ge N\ve$ whp. By the upper bound in Eq.~(\ref{eq:zz5})
and monotonicit of $\Xi(\zeta)$, we get
\begin{eqnarray}
\log\Xi(N\ve)\le \log Z(\beta_1)+2\beta_1 U(\beta_1)\le
 \log Z(\beta_1)+2\beta_1 N C_2e^{-b\beta_1}+2\beta_1C_3N^{\delta}\, .
\end{eqnarray}
Using the definition of $\beta_1$, which gives 
$\beta_1 = \frac{1}{2}\log\frac{2C_1}{\ve}$, we get that there exists 
$C, a>0$ such that
\begin{eqnarray}
\log\Xi(N\ve)\le \log Z(\beta_1)+NC\ve^{a}\, .
\end{eqnarray}
with high probability.

The lower bound on $\log\Xi(N\ve)$ is proved analogously by taking
$\beta_2$ such that  $2C_2e^{-b\beta}+2C_3N^{-1+\delta}=\ve$
thus getting $\log\Xi(N\ve)\ge \log Z(\beta_2)-NC\ve^{a}$ whp.
One concludes by bounding the difference of the two partition 
functions:
$|\log Z(\beta_2)-\log Z(\beta_1)|\le U(\beta_2)|\beta_1-\beta_2|\le 
NC\ve^{a}$ whp.
\endproof

\begin{thebibliography}{99}

\bibitem{Georgii} H.~O.~Georgii,
``Gibbs Measures and Phase Transitions''. 
Berlin, Walter de Gruyter and Co., 1988

\bibitem{Monasson} R.~Monasson, R.~Zecchina, S.~Kirkpatrick, B.~Selman,
and L.~Troyansky, ``Determining  computational complexity
from characteristic `phase transitions' '',
Nature 300 (1999), 133--137

\bibitem{Mezard} M.~M\'ezard, G.~Parisi, and R.~Zecchina,
``Analytic and Algorithmic Solution of Random Satisfiability Problems'',
Science 297 (2002), 812--815

\bibitem{ANPNature} D.~Achlioptas, A.~Naor and Y.~Peres,
``Rigorous location of phase transitions in hard optimization problems'',
Nature 435 (2005), 759--764

\bibitem{T01} M.~Talagrand, 
`` The high temperature case of the K-sat problem", 
Probability Theory and Related Fields 119, 2001, 187-212.

\bibitem{AP04} D. Achlioptas and Y. Peres, 
``The threshold for random $k$-SAT is $2k  \log 2 - O(k)$",
Journal of the AMS, 17 (2004), 947-973.

\bibitem{MonassonZecchina} R.~Monasson and R.~Zecchina,
``Entropy of the $K$-Satisfiability Problem'',
 Phys. Rev. Lett. 76 (1996), 3881–3885 
%
%\bibitem{ANP05} D. Achlioptas, A. Naor and Y. Peres, ``The Fraction of Satisfiable 
%Clauses in a Random Formula", Preliminary version in Proceedings of IEEE FOCS 2003. Long
%version to appear in the Journal of ACM.

\bibitem{FL03} S.~Franz and M.~Leone, 
``Replica bounds for optimization problems and diluted spin systems",
Journal of Statistical Physics,  111 (2003), 535.

\bibitem{FLT03} S.~Franz, M.~Leone and F.~L.~Toninelli, 
``Replica bounds for diluted  non-Poissonian spin systems", 
Journal of Physics, A 36 (2003) 10967 . 

\bibitem{JS93}  M.~Jerrum and A.~Sinclair, 
``Polynomial-time Approximation  Algorithms for the Ising Model", 
SIAM Journal on Computing 22 (1993), pp. 1087-1116.

\bibitem{W06} D.~Weitz, 
``Counting independent sets up to the tree threshold'', 
In Proceedings of STOC, 2006.

\bibitem{BG06} A.~Bandyopadhyay and D.~Gamarnik, ``Counting without sampling. New algorithms for 
enumeration problems using statistical physics'', In Proceedings of SODA, 2006.

\bibitem{Future} A.~Montanari and D.~Shah, 
``$k$-SAT: Counting Satisfying Assignment and Threshold for Correlation Decay",
Longer version, in preparation.

\bibitem{Friedgut} E.~Friedgut, 
``Sharp Thresholds of Graph Proprties, and the $k$-sat Problem'',   
Journal of American Mathematical Society, 12 (1999), no. 4, 1017--1054.

\bibitem{Pearl} J.~Pearl, 
``Probabilistic Reasoning in Intelligent Systems: Networks
of Plausible Inference'', 
San Francisco, CA: Morgan Kaufmann, 1988.

\bibitem{WJ} M.~Wainwright and M.~Jordan, 
``Graphical models, exponential families,
and variational inference,'' \emph{Tech.\ Report}, Dept. of
Stat.,University of Cal., Berkeley, 2003. 

\bibitem{TJ99} S.~Tatikonda and M.~Jordan, ``Loopy Belief
Propagation and Gibbs Measure,'' Berkeley Working Paper, 2002.

\end{thebibliography}
\end{document}